\documentclass{article}
\usepackage[accepted]{vietnam} 
\usepackage{natbib}
\usepackage{graphicx, epsf}       
\usepackage[breaklinks,colorlinks,citecolor=blue,linkcolor=magenta]{hyperref}
\usepackage{natbib}
\usepackage{color}
\usepackage{multirow}
\usepackage{appendix}
\newcommand{\Msun}{$M_\odot$}
\newcommand{\Mstar}{$M_{\star}$}
\newcommand{\Mbh}{$M_{\rm BH}$}

\renewcommand{\deg}{\ensuremath{^{\circ}}}
\newcommand{\ml}{\emph{M/L}}
\newcommand{\mleff}{\emph{M/L}$_{\rm eff}$}

\newcommand{\hst}{\emph{HST}}
\newcommand{\FI}{F814W}
\newcommand{\FB}{F336W}

\begin{document}
\twocolumn[
\title{Improved dynamical constraints on the mass of the central black hole in NGC~404}
\titlerunning{Improved dynamical constraints on the mass of the central black hole in NGC~404}
\author{Dieu D. Nguyen}{dieu.nguyen@utah.edu}
\address{Department of Physics and Astronomy, University of Utah, 115 South 1400 East, Salt Lake City, UT 84112, USA}

\vskip 0.5cm 
]
%====================================================================%
% Abstract section!!!
%====================================================================%
\begin{abstract}
We determine the dynamical black hole mass in NGC~404 including modeling of the nuclear stellar populations.  We combine \hst/STIS spectroscopy with WFC3 images to create a local color--\ml~relation derived from stellar population modeling of the STIS data.  We then use this to create a mass model for the nuclear region. We use Jeans modeling to fit this mass model to adaptive optics stellar kinematic observations from Gemini/NIFS.  From our stellar dynamical modeling, we find a  3$\sigma$ upper limit on the black hole mass of $1.5\times10^5$\Msun.
\end{abstract}

%====================================================================%
% Introduction Section!!!
%====================================================================%

\section{Introduction}
The nearby galaxy NGC~404 provides an excellent opportunity to study the relationships of NSCs and MBHs.   NGC~404 is a dwarf S0 galaxy \citep[\Mstar~$\sim10^9$\Msun;][]{Seth10} at a distance of just $\sim$3.0 Mpc (Karachentsev et al., 2002).    The galaxy appears to host an MBH within its prominent central nuclear star cluster \citep[NSC;][]{Seth10}.   Dynamical measurements using high-resolution adaptive optics data from Gemini/NIFS by \citep{Seth10} indicate the possible detection of an MBH, with a firm upper limit of \textless$10^6$\Msun.  Their dynamical modeling depended on a number assumptions, most importantly the assumption of a constant \ml~throughout the nucleus.  This is a poor assumption due to the obvious spatial gradients of stellar populations within the nucleus.  The stellar mass in the central resolution element of the Gemini/NIFS data was comparable to the BH mass (i.e.,~the sphere of influence was not well resolved), hence, any spatial gradients in the \ml~can have a significant effect on the BH mass estimate.  In this work, we use new STIS spectroscopy and WFC3 imaging to quantify the spatial variations in the \ml~throughout the NGC~404 nucleus and improve the BH mass estimate.  Incorporating \ml~gradients to refine BH mass estimates was recently explored by \citet{McConnell13a}, who used color gradients to explore possible radial \ml~gradients in three giant elliptical galaxies.

%====================================================================%
% Color--M/L Correlation and Mass Model !!!
%====================================================================%
\section{Color--\emph{M/L} Correlation} 
%%%%%%% Figure 1 %%%%%%%%%%%%%%%
\begin{figure*}[!htb]
\minipage{0.5\textwidth}
  \includegraphics[width=\linewidth]{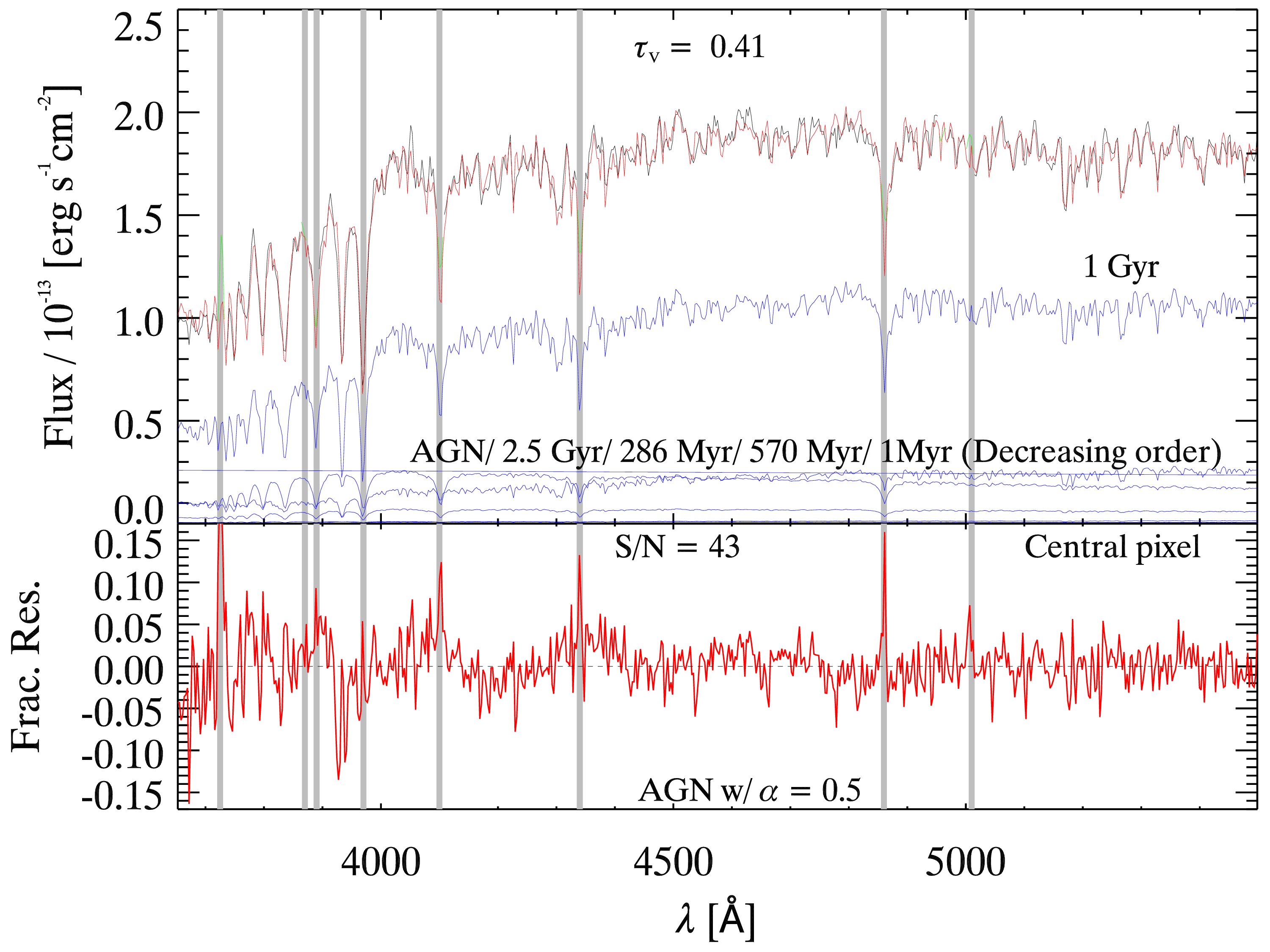}\label{ssp_agn_fit}
\endminipage\hfill
\minipage{0.5\textwidth}
  \includegraphics[width=\linewidth]{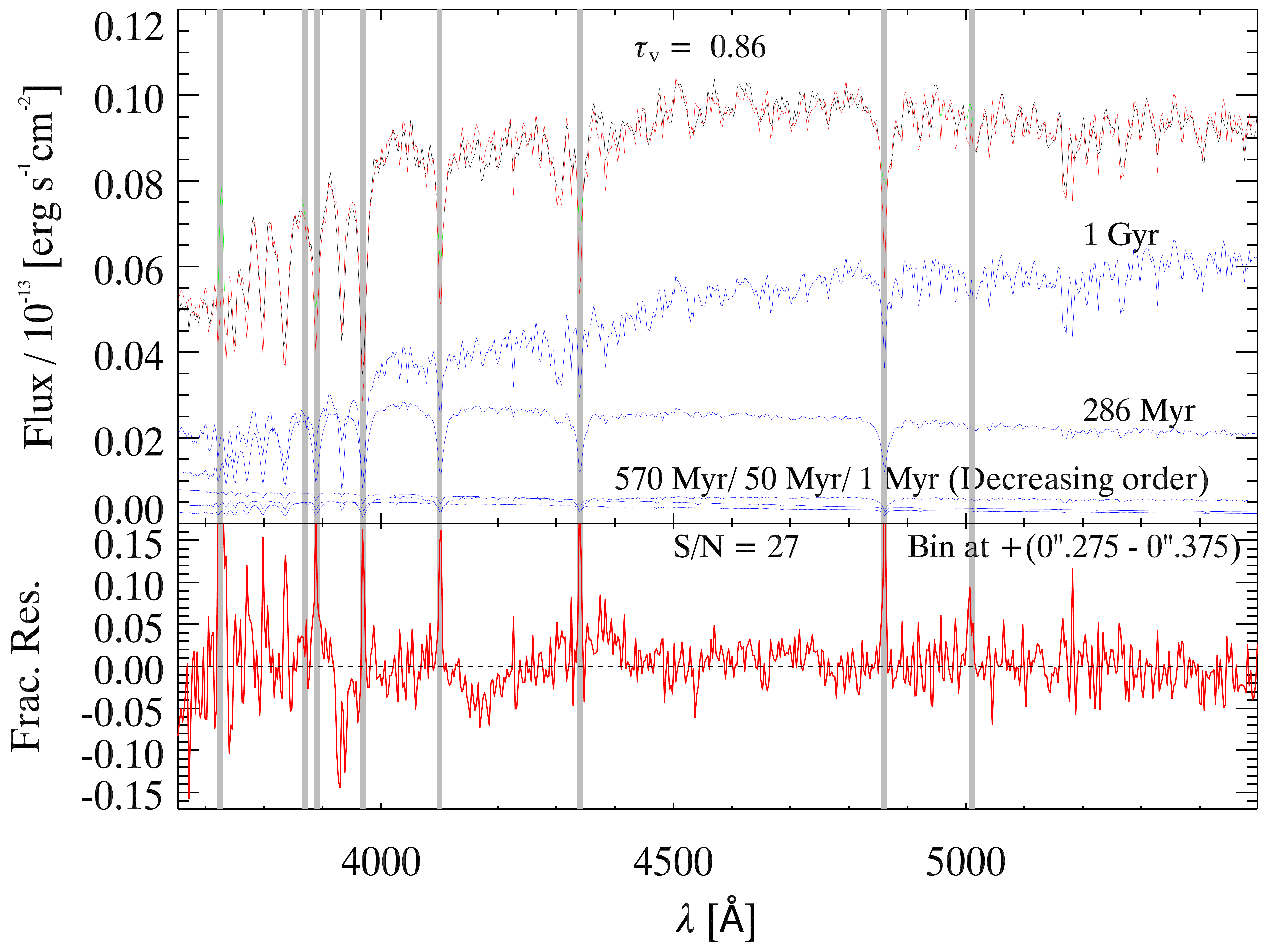}\label{ssp_noagn_fit}
\endminipage
\caption[SCP06C1]{Left top-panel: best-fit stellar population models to individual pixel spectra at the center. Right top-panel: best-fit stellar population models to a binned spectrum at large radius spanning radii of $+($0\farcs275--0\farcs375).  The central spectrum  is fitted with an AGN continuum component, while the outer bin is not.  The STIS spectrum is shown in black, the best-fitting stellar population synthesis model fits \emph{including an AGN component} with $\alpha=0.5$ shown in red.  Vertical grey lines show emission line regions that were excluded from the fit. Blue spectra indicate the different ages SSPs and AGN component contributing to the multi-age fit.   Right bottom-panels: fractional residuals of the best-fit multi-age  spectrum.}
\label{ssp_fit}
\end{figure*}
%%%%%%% %%%%%%%%%%%%%%%%%%%%%%%%%%%%% %%%%%%%%%%%%%%%%%%%%%%

%%%%%%% Figure 2 %%%%%%%%%%%%%%%%%%%%%% %%%%%%%%%%%%%%%%%%%%%%
\begin{figure*}[tbh]
\begin{minipage}[b]{3.0in}
\includegraphics[scale=0.14]{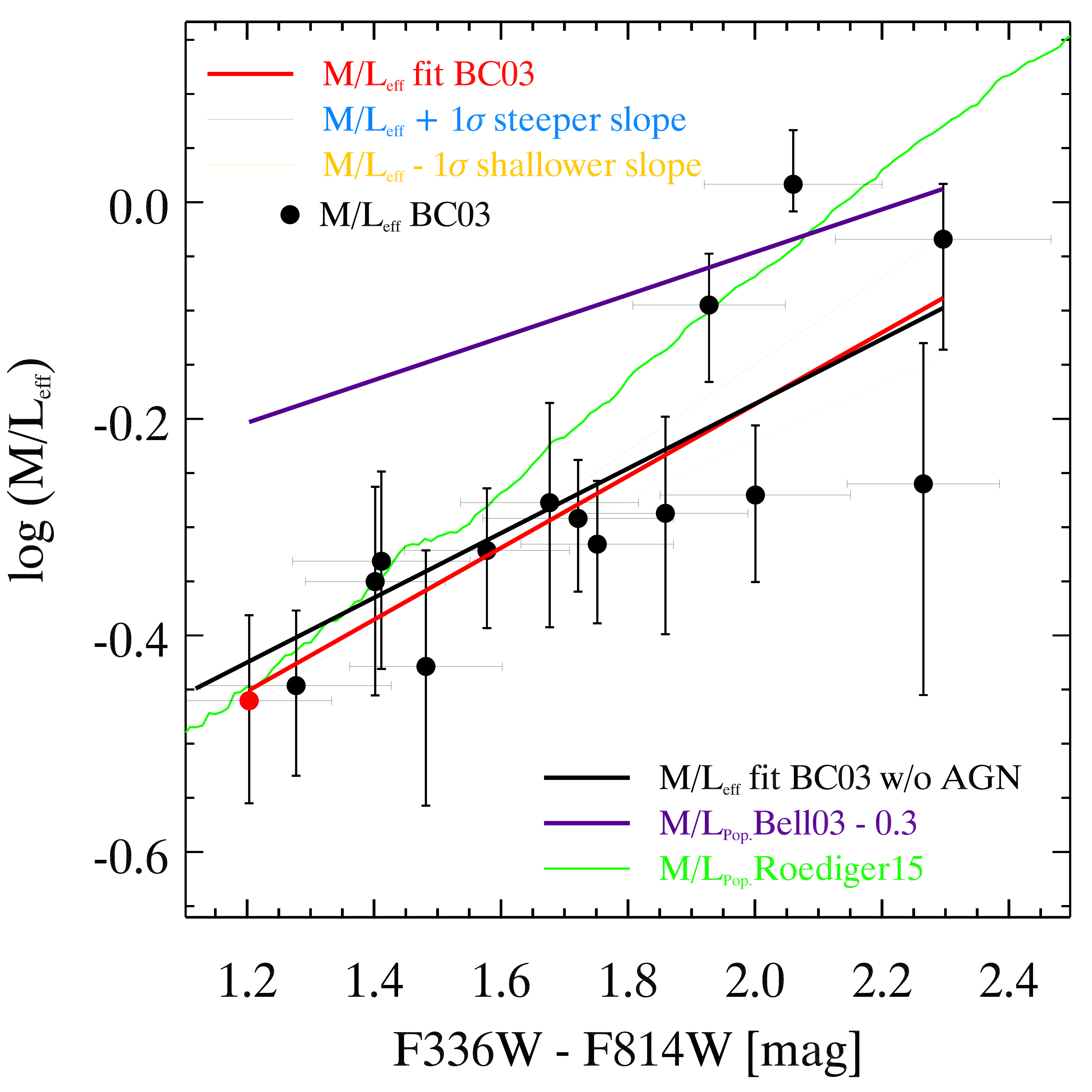}
\end{minipage}
\vspace{-5mm}
\begin{minipage}[b]{3.7in}
\caption{The mass-to-light ratio -- color relation for the NGC~404 nucleus.  The y-axis shows the effective mass-to-light ratio in \FI~(\mleff) determined from stellar population fits to the STIS spectroscopy, while the x-axis shows the \FB--\FI~color determined from WFC3 imaging.  Black points show the data for the stellar population fits using the BC03 models, while the thick red line shows the best-fit linear relation to these data. The 1$\sigma$ uncertainties in the slope of our best-fit linear relation are shown with thin blue (steeper) and yellow (shallower) lines.  The black thin line shows the best-fit linear relation to the data when we don't include an AGN component in the five innermost spectral bins during the fitting procedure. Error bars for each point in log(\mleff) were determined through a Monte Carlo analysis of the stellar population fits and these errors form the dominant error in our best-fit log(\mleff) color relations. The purple thick solid line is the predicted color--\ml~correlation from \citet{Bell03} shifted downwards by 0.3 dex, while the green line shows the \citet{Roediger15} relation.  The red data point indicates the NGC 404 center.}
\label{color_m2leff_correlation}
\end{minipage}
\end{figure*}
%%%%%%% %%%%%%%%%%%%%%%%%%%%%%%%%%%%% %%%%%%%%%%%%%%%%%%%%%%

We fit a range of single stellar population (SSP) models to the nuclear STIS spectroscopic data to determine the ages of stars and~\ml~spatial variations within the NGC~404 NSC. We show that the nuclear spectrum is somewhat better fit with the addition of an AGN continuum component, and use this as our default model in the nuclear regions.  The best-fit model and residuals to the central pixel are illustrated in the left panel of Figure~\ref{ssp_fit}.   The model is shown in red with the individual SSP components shown in blue.     The fit in the central pixel is dominated by intermediate age stars with 70\% of the light in the 1 Gyr SSP, 5\% of the light in the 2.5 and 5~Gyr SSP, 17\% of the light in AGN. Only a small fraction of the light is in younger populations (\textless570 Myr).   The right panel of this figure shows a lower S/N spectral bin spanning $+(0\farcs275-0\farcs375)$ with higher extinction.  The reduced $\chi^2$ is 0.98 for the central pixel and 2.23 for the outer bin; the decrease in the goodness-of-fit at larger radii is likely due to increased contributions from bad pixels.

Figure~\ref{color_m2leff_correlation} shows the correlation of the WFC3 \FB--\FI~color map with the spectroscopic~\mleff~values. We fit this correlation to get a color--\ml~relation; the best-fit is the red solid line.  The idea behind this approach is the same as in \citet{Bell03} and \citet{Zibetti09}, but appropriate to the exact populations present in the NGC~404 nucleus as determined using the STIS population synthesis fits.  We fit a linear relation between the logarithm effective spectral~\mleff~and the \FB--\FI~color. The error on this relation is determined in two ways: first, we propagate the errors on the spectroscopic~\ml~determined above, and second, we determine bootstrap errors of the fit by refitting using random sampling with replacement. The first method yields significantly larger errors than our bootstrapped errors, which are minimal.   We calculate the 1$\sigma$ uncertainty on our color--\ml~relation from our total error budget and show these as the thin blue (steeper) and yellow (shallower) lines in Figure~\ref{color_m2leff_correlation}, respectively. We will examine the effect of our color--\ml~relation uncertainties on the dynamical models in Section~\ref{sec:jeans}.

%====================================================================%
% STELLAR-DYNAMICAL MODELING
%====================================================================%
%%%%%%% Figure 3 %%%%%%%%%%%%%%%%%%%%%% %%%%%%%%%%%%%%%%%%%%%%
\begin{figure*}[!h]
\vspace{5mm}
\minipage{0.33\textwidth} 
  \includegraphics[width=\linewidth]{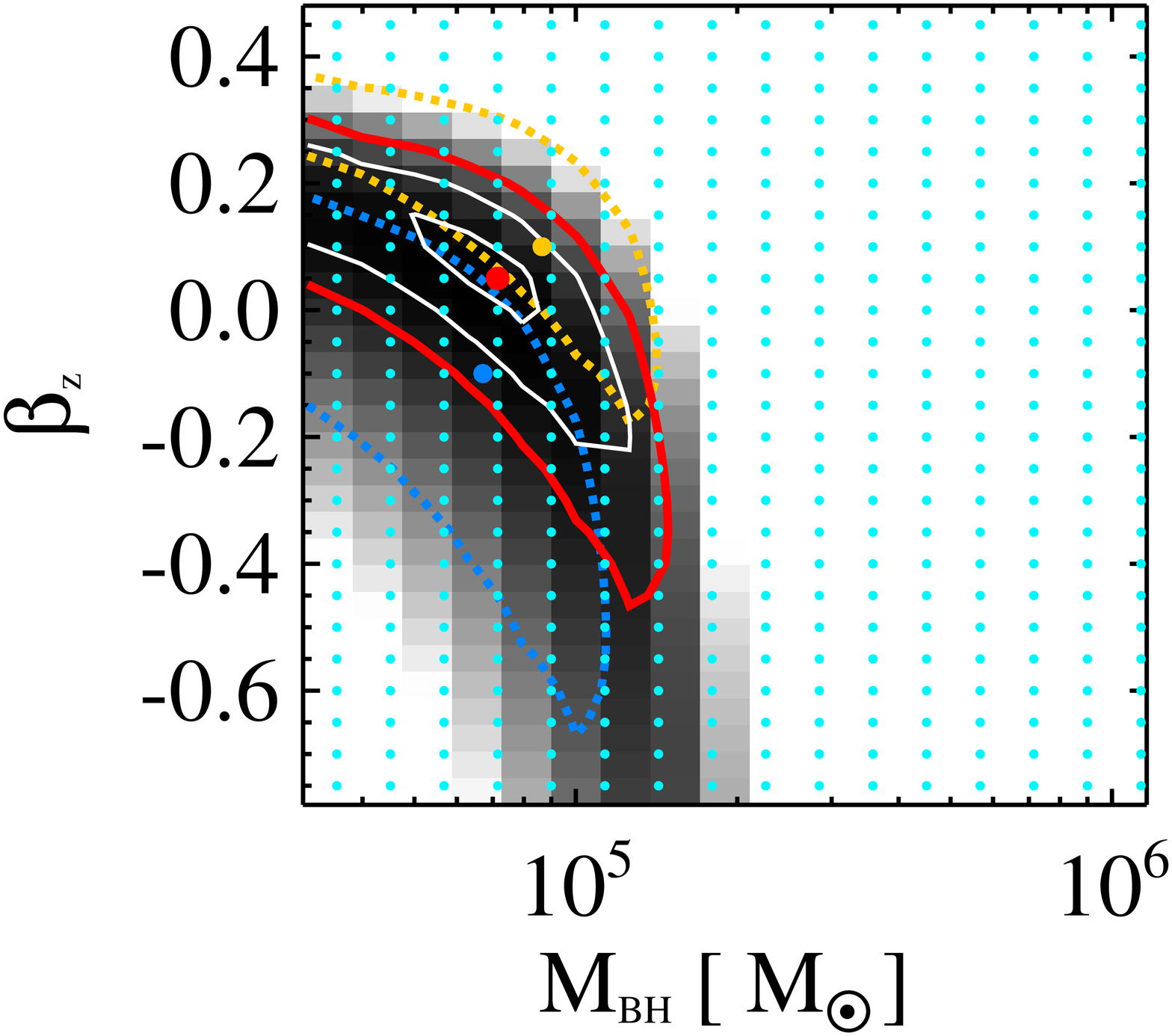}\label{jam_betaz_mbhs}
\endminipage\hfill
\minipage{0.33\textwidth}
  \includegraphics[width=\linewidth]{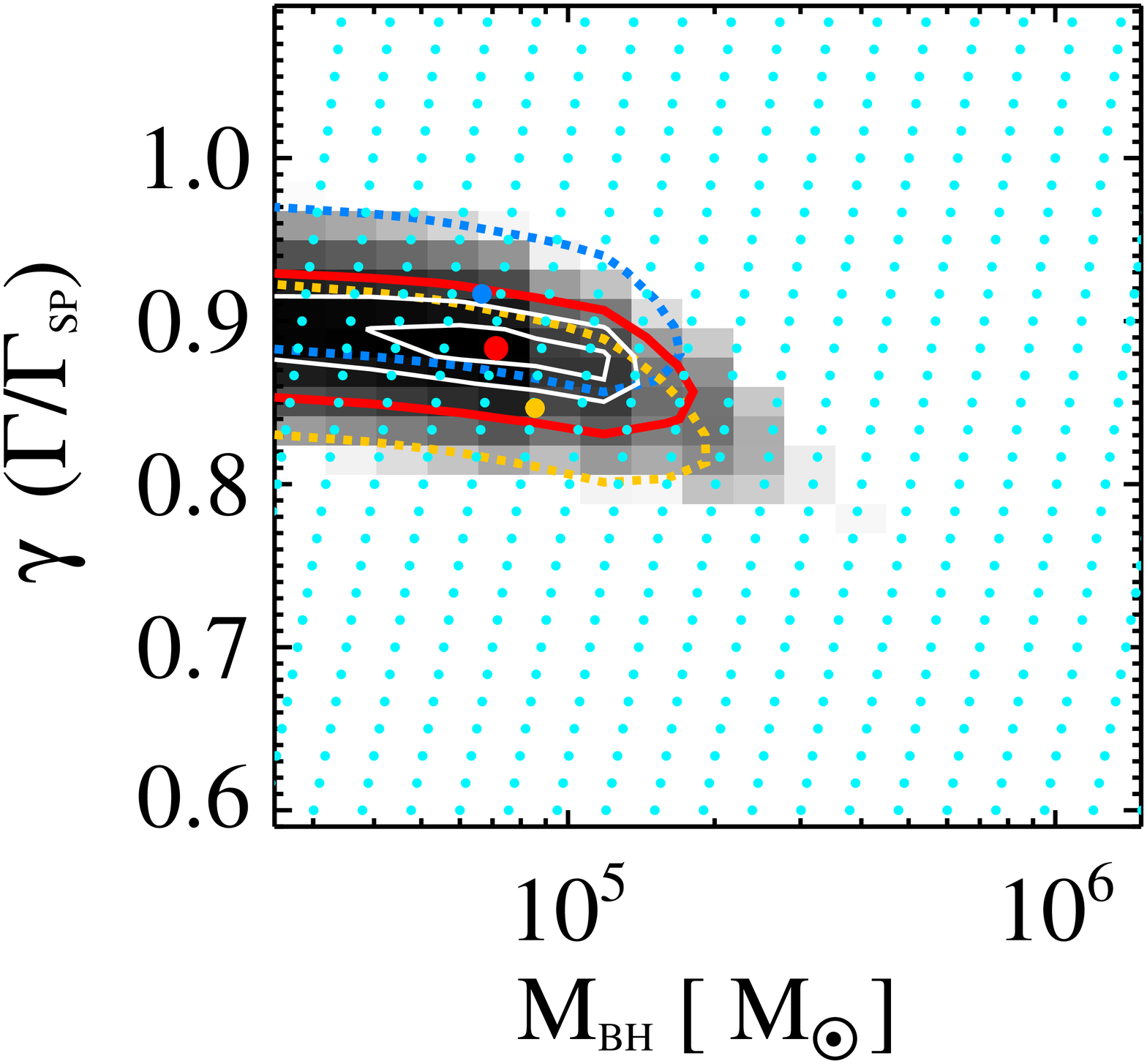}\label{jam_scale_mbhs}
\endminipage\hfill
\minipage{0.33\textwidth}
  \includegraphics[width=\linewidth]{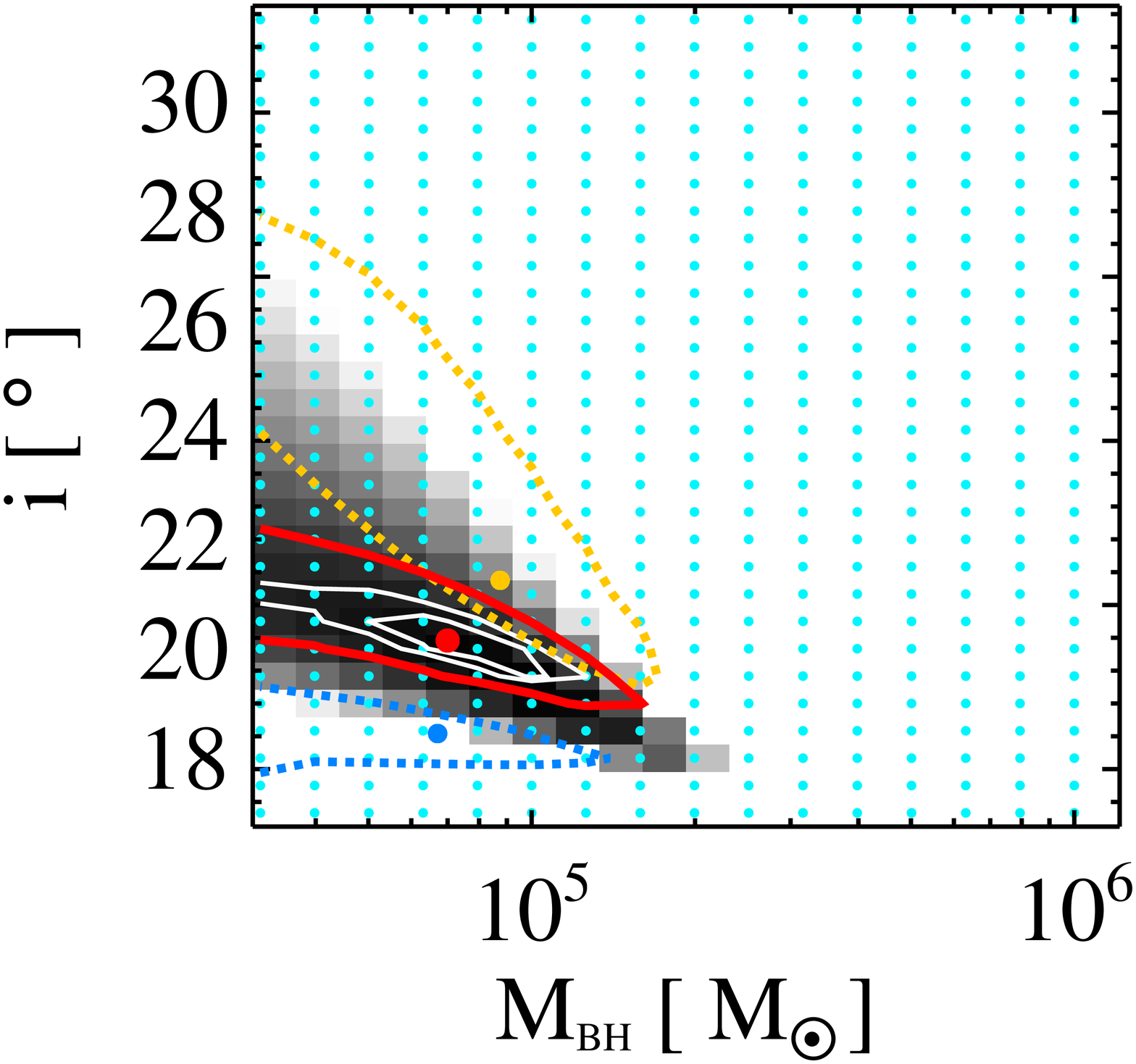}\label{jam_incl_mbhs}
\endminipage
\caption[SCP06C1]{ Best-fit JAM models using the Gemini/NIFS $V_{\rm rms}$ data and our updated mass map MGE model. The models optimize the four parameters \Mbh, $\beta_z$, $\gamma$ (the ratio between the dynamical mass and the stellar population mass), and $i$.  The cyan dots show the grid of models in  each panel. The grayscale shows the likelihood of the best-fit model.  {\em Left panel:} the best-fit anisotropy ($\beta_z$) vs.~M$_{\rm BH}$. The red dot shows the minimum $\chi^2$.  $\chi^2$ contours are shown at $\Delta\chi^2 = 2.30, 6.18$ (white solid lines), ${\rm and \;} 11.83$ (red solid line) corresponding to 1$\sigma$, 2$\sigma$, and 3$\sigma$ confidence levels for two parameters after marginalizing over the other two parameters.   Similarly, the blue and yellow dashed lines illustrate the $\chi^2$ contours of the 3$\sigma$ confidence levels for the mass maps created from the steeper and shallower color--\ml~slope mass map models (corresponding to the thin blue and yellow lines show in Figure~\ref{color_m2leff_correlation}).  {\em Middle panel:} the best-fit mass scaling factor, $\gamma$ vs.~\Mbh; markings as in left panel. {\em Right panel:} the inclination angle $i$ of the galaxy vs.~\Mbh.}
\label{jam_models}
\end{figure*}  
%%%%%%%%%%%%%%%%%%%%%%%%%%%%%%%%%% %%%%%%%%%%%%%%%%%%%%%%

\section{Stellar Dynamical Modeling}\label{sec:jeans} 
We use the Jeans Anisotropic Modeling (JAM) method and \texttt{IDL} software as the dynamical model to calculate the mass of central BH, \Mbh. The model relies on an axisymmetric solution of the Jeans equations incorporating orbital anisotropy \citep{Cappellari08}. The anisotropy is characterized by a single anisotropy parameter: $\beta_z=1-\sigma_z^2/\sigma_R^2$, where $\sigma_R$ and $\sigma_z$ are the velocity dispersion in the radial direction and z-direction in ellipsoid aligned cylindrical coordinates. The model calculates the predicted second velocity moment, $V_{\rm rms}=\sqrt{V^2+\sigma^2}$, where $V$ is the mean stellar velocity and $\sigma$ is the velocity dispersion based on an MGE model.   We note we are using a mass model that differs from the luminosity profile of the galaxy; this difference is incorporated into the JAM model by using separate luminosity and mass profile MGEs; however both are axisymmetric and thus may not fully capture the variations in kinematics due to e.g.,~dusty regions.  We also note that because of the proximity of NGC~404, the kinematic maps have some contribution from partially resolved stars that make the dynamical maps appear less smooth, especially at larger radii. The model has four parameters: (1) the mass of a central BH, \Mbh, (2) the mass scaling factor $\gamma$; while for a normal JAM fit, this would be the dynamical \ml, in our case this parameter is the ratio between the dynamical \ml~and the \ml~predicted by our stellar population fitting (which assumes a Chabrier IMF), (3) the anisotropy, $\beta_z$, and (4) the inclination angle, $i$.

We compute the models on a grid of $\beta_z$, \Mbh, $\gamma$, and $i$. At each grid point we evaluate the $\chi^2$ of the the predicted $V_{\rm rms}$ relative to the kinematic data.   We run the dynamical modeling to fit these four parameters ($\beta_z$, \Mbh, $\gamma$, $i$) in 2--D using 920 kinematic data points (degrees of freedom, dof) within a radius of $\sim1\farcs5$ of the center. Figure~\ref{jam_models} shows $\chi^2$ contours as a function of \Mbh~vs. the anisotropy parameter $\beta_z$ (left), scaling factor $\gamma$ (middle), and inclination angle $i$ (right) after marginalizing over the other two parameters.  The minimum reduced $\chi^2_r\sim1.26$~is found at \Mbh~= $7.0\times10^4$\Msun, $\beta_z\sim0.05$ (i.e., nearly isotropic), $\gamma=0.890$, and $i=20$\deg.  The solid contours show the  $1\sigma$, $2\sigma$ (white), and $3\sigma$ (red) levels (or $\Delta\chi^2=2.30$, $6.18$, and $11.83$) for two parameters; we choose the confidence limits for two parameters to accurately capture the uncertainties shown in our two parameter plots above after marginalizing over the third and the fourth parameter in each plot.  The BH is only detected at the 1$\sigma$ level.  Given the restrictions JAM models place on the orbital freedom of the system, it is common to use 3$\sigma$ limits in quoting BH masses.  Therefore, we use the 3$\sigma$ upper limit of \Mbh~$<1.5\times10^5$\Msun, one order of magnitude lower than \citep{Seth10} estimate.  

The most significant change in comparison with \citep{Seth10} is the anisotropy parameter, $\beta_z$.  Our best model is more isotropic ($\beta_z=0.05^{+0.05}_{-0.15}$) than the best fit of $\beta_z=0.5$ found in \citet{Seth10}.  Given the observed isotropy in other galaxy nuclei \citep{Seth14}, we interpret this as a sign of the success of our mass model (which incorporates variations in~\mleff) correctly representing the mass distribution in this system.

%====================================================================%
% Summary
%====================================================================%
\section{Summary}
We develop a method to incorporate variations in the stellar~\ml~into the dynamical modeling to constrain the BH mass in NGC~404.  Specifically, we use spectroscopically determined~\ml~spatial variations of the nuclear region and use these to create~\ml~vs. color relations appropriate to the local stellar populations present in the nucleus.  We then use these relations to create a mass map of the nucleus.  The spatial variabilities in~\ml~are thus directly incorporated into our dynamical model. Incorporating stellar population-based models are critical for getting good dynamical constraints on the lowest mass BHs, as most of these are located in NSCs with complicated and spatially varying stellar populations.  
 
Jeans anisotropic models of the stellar kinematics with the derived mass map gives a BH mass of \Mbh~$=7.0^{+1.5}_{-2.0}\times10^4$\Msun, a mass scaling factor (the ratio of the dynamical to stellar population mass) of $\gamma=0.890_{-0.060}^{+0.045}$, an anisotropy parameter $\beta_z=0.05^{+0.05}_{-0.15}$, and an inclination angle $i=20.5^{+1.0}_{-2.0}$.  The BH mass is consistent with zero at the 3$\sigma$ level, thus, we present a 3$\sigma$ upper limit to the mass of $1.5\times10^5$\Msun.  This BH mass upper limit suggests NGC~404 falls clearly below scaling relations between the BH mass and the bulge or total mass while it is consistent with the $M-\sigma$ relation.  This is the lowest dynamical BH mass estimate ever performed via the stellar dynamical method.

%====================================================================%
% acknowledgments
%====================================================================%
%\acknowledgments
%The authors would like to thank the University of Utah, Physics and Astronomy Department, for supporting this work based on the financial support from~\hst~grant GO-12611 and from NSF grant AST-1350389.
%====================================================================%
% references 
%====================================================================%
%\tiny
%\bibliography{dnguyenref}

\begin{thebibliography}
\bibitem[{{Bell} {et~al.}(2003){Bell}, {McIntosh}, {Katz}, \&
  {Weinberg}}]{Bell03}
{Bell}, E.~F., {McIntosh}, D.~H., {Katz}, N., \& {Weinberg}, M.~D. 2003, \apjs,
  149, 289

\bibitem[{{Cappellari}(2008)}]{Cappellari08}
{Cappellari}, M. 2008, \mnras, 390, 71

\bibitem[{{McConnell} {et~al.}(2013){McConnell}, {Chen}, {Ma}, {Greene},
  {Lauer}, \& {Gebhardt}}]{McConnell13a}
{McConnell}, N.~J., {Chen}, S.-F.~S., {Ma}, C.-P., {et~al.} 2013, \apjl, 768,
  L21

\bibitem[{{Roediger} \& {Courteau}(2015)}]{Roediger15}
{Roediger}, J.~C., \& {Courteau}, S. 2015, \mnras, 452, 3209

\bibitem[{{Seth} {et~al.}(2010){Seth}, {Cappellari}, {Neumayer}, {Caldwell},
  {Bastian}, {Olsen}, {Blum}, {Debattista}, \citep{} {McDermid}, {Puzia}, \&
  {Stephens}}]{Seth10}
{Seth}, A.~C., {Cappellari}, M., {Neumayer}, N., {et~al.} 2010, \apj, 714, 713

\bibitem[{{Seth} {et~al.}(2014){Seth}, {van den Bosch}, {Mieske}, {Baumgardt},
  {Brok}, {Strader}, {Neumayer}, {Chilingarian}, {Hilker}, {McDermid},
  {Spitler}, {Brodie}, {Frank}, \& {Walsh}}]{Seth14}
{Seth}, A.~C., {van den Bosch}, R., {Mieske}, S., {et~al.} 2014, \nat, 513, 398

\bibitem[{{Zibetti} {et~al.}(2009){Zibetti}, {Charlot}, \& {Rix}}]{Zibetti09}
{Zibetti}, S., {Charlot}, S., \& {Rix}, H.-W. 2009, \mnras, 400, 1181

\end{thebibliography}

\end{document}